\documentclass{appolb}
\usepackage{graphicx}

\usepackage{amsmath}
\usepackage{amssymb}
\usepackage{graphicx}
\usepackage{dsfont}
\usepackage{dcolumn}
\usepackage{units}
\usepackage{bm}
\usepackage{wasysym}
\usepackage{multirow}
\usepackage{slashed}
\usepackage[usenames]{color}
\usepackage[colorlinks=true,linkcolor=blue,citecolor=blue,urlcolor=blue]{hyperref}
\usepackage{cite}

      \newcommand{\Slash}[1]{\slashed{#1}}

\begin{document}


\title{Probing nucleons with photons at the quark level
\thanks{Presented at the Workshop "Excited QCD 2014", Bjela\v snica Mountain, Sarajevo, Bosnia-Herzegovina, February 2-8, 2014}
}

\author{Gernot Eichmann
\address{Institut f\"{u}r Theoretische Physik, Justus-Liebig-Universit\"{a}t Giessen, \\ 35392 Giessen, Germany}}

       \maketitle
       \vspace{-3mm}
       \begin{abstract}
       The description of electromagnetic interactions with hadrons from the quark level requires knowledge of the
       underlying quark-gluon ingredients. I discuss some properties of the quark-photon vertex and
       quark Compton vertex, along with the role of electromagnetic gauge invariance and vector-meson dominance.
       A simple parametrization for the quark-photon vertex is given.
       \end{abstract}


       \section{Introduction}

       The electromagnetic interaction plays an important role in mapping out the internal quark-gluon structure of the nucleon.
       This is evidenced by a sizeable number of ongoing and future
       experiments at Jefferson Lab, MAMI, ELSA, or the upcoming PANDA/FAIR experiment.
       Among the various theoretical approaches, also the Dyson-Schwinger equations (DSEs) of QCD~\cite{Roberts:1994dr,Alkofer:2000wg,Fischer:2006ub} have provided 
       insight along the way. Already a rainbow-ladder truncation, where
       quarks and gluons interact through a tree-level vertex only, and quarks and antiquarks via gluon exchange, has proven quite useful in describing a range of hadron properties.
       These include pseudoscalar and vector-meson spectra, their form factors and other structure properties~\cite{Bashir:2012fs};
       but also nucleon and $\Delta-$baryon observables such as masses and electromagnetic,
       axial and transition form factors~\cite{Eichmann:2009qa,Eichmann:2011vu,Eichmann:2011pv,Mader:2011zf,Eichmann:2011aa,Sanchis-Alepuz:2013iia}.
       The main missing contributions to form factors presumably come from the cloud of virtual pions that 'dress' the nucleon.
       Calculations beyond rainbow-ladder to shed some light on these issues are underway~\cite{Chang:2009zb,Fischer:2009jm,Sanchis-Alepuz:2014wea,Williams:2014iea}.

       The question is then: why does such a 'simple' truncation work at all?
       Perhaps one answer is that ground-state hadrons (the '$s$ waves' in the quark model)
       are not very sensitive to the details of the quark-gluon interaction.
       It is then scalar and axialvector mesons, heavy-light systems or excited hadrons where mismatches should appear (and they do indeed~\cite{Krassnigg:2009zh,Nguyen:2010yh}).
       \textit{Symmetries} are another answer: they are implemented at the quark level and carefully maintained throughout every step in these calculations.
       Through vector and axialvector Ward-Takahashi identities (WTIs), a gluon ladder kernel is linked to the gluon exchange that defines the quark DSE.
       One cannot simply add interaction diagrams in a $q\bar{q}$ system without dressing the quark-gluon vertex simultaneously, and all other Green functions will undergo changes as well.
       These symmetries ensure that the pion is a $q\bar{q}$ bound state but also QCD's Goldstone boson in the chiral limit.
       They enforce electromagnetic current conservation for electromagnetic form factors, the Goldberger-Treiman relation for axial form factors and so on, so that no 'fine-tuning' is necessary.

       In order to calculate nucleon form factors and polarizabilities, we must couple photons to nucleons in a symmetry-preserving way~\cite{Kvinikhidze:1999xp,Oettel:1999gc,Eichmann:2011ec}.
       To this end, we should first understand how a photon microscopically interacts with a quark.
       Two of the relevant Green functions that encode this interaction are the quark-photon vertex and the quark Compton vertex.
       Here I will discuss some of their properties, the role of electromagnetic gauge invariance in determining their structure,
       and their implications for hadron properties.

            \begin{figure}[t!]
            \centerline{%
            \includegraphics[width=14cm]{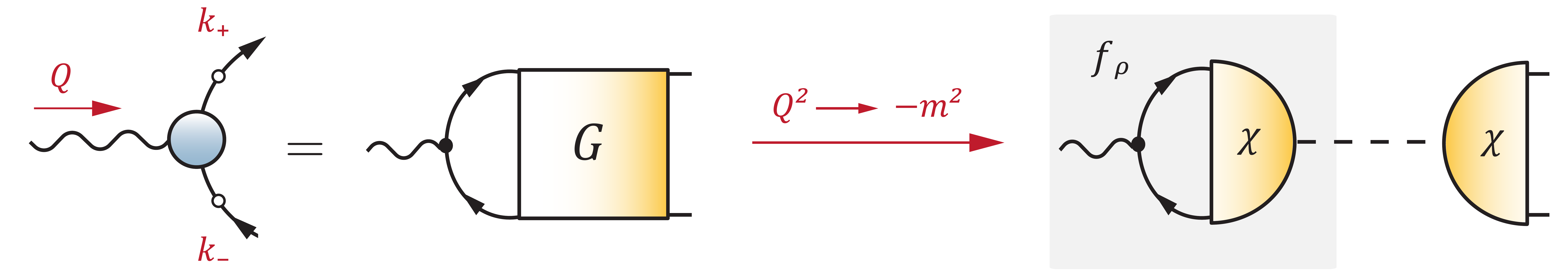}}
            \caption{Quark-photon vertex and the $\rho-$meson poles it contains.}
            \label{fig:qpv}
            \end{figure}

       \section{Quark-photon vertex}

             Several well-known characteristics of form factors are reflected in the nonperturbative structure of the dressed quark-photon vertex.
             The vertex is defined as the $\gamma^\mu-$contraction of the $q\bar{q}$ four-point function, see Fig.~\ref{fig:qpv}.
             The four-point function contains all intermediate hadronic states that can be formed by a valence quark and antiquark.
             Therefore, its singularity structure in the vector channel will be inherited by the quark-photon vertex, i.e.,
             'vector-meson dominance' is implemented by construction.
             On the other hand, the definition allows to derive an inhomogeneous Bethe-Salpeter equation (BSE) for the vertex; it depends on the $q\bar{q}$ kernel where the truncation to rainbow-ladder is made.
             Its numerical solution has been first achieved in Ref.~\cite{Maris:1999bh} and nowadays become almost a routine task.
             However, even before solving the vertex dynamically one can gain some insight based on general properties alone.

             Electromagnetic gauge invariance entails that the quark-photon vertex can be separated into a 'gauge part' and a purely transverse part:
                  \begin{equation}\label{vertex:BC}
                      \Gamma^\mu(k,Q) = \Big[  i\gamma^\mu\,\Sigma_A + 2 k^\mu (i\Slash{k}\, \Delta_A  + \Delta_B) \Big] +  \Big[ i\sum_{j=1}^8 f_j \, \tau_j^\mu(k,Q) \Big].
                  \end{equation}
             Here, $Q$ is the photon momentum and $k=(k_++k_-)/2$ the average momentum of the quark legs, see Fig.~\ref{fig:qpv}.
             The gauge part in the first bracket is the Ball-Chiu vertex~\cite{Ball:1980ay} that satisfies the vector WTI.
             It is completely determined by the dressed fermion propagator.
             At large $Q^2$ it reproduces the tree-level structure, whereas the nonperturbative dressing effects are contained in $\Sigma_A$, $\Delta_A$ and $\Sigma_B$.
             These are sums and difference quotients of the quark dressing functions $A(p^2)$ and $B(p^2)$:
                 \begin{equation}\label{QPV:sigma,delta}
                     \Sigma_F(k,Q) = \frac{F(k_+^2)+F(k_-^2)}{2} , \qquad
                     \Delta_F(k,Q) = \frac{F(k_+^2)-F(k_-^2)}{k_+^2-k_-^2},
                 \end{equation}
              with $F\in\{A,B\}$. $A(p^2)$ approaches the quark wave-function renormalization constant $Z_2$ at large $p^2$ and is nonperturbatively enhanced.
              The quark mass function $M(p^2)=B(p^2)/A(p^2)$ passes through the current-quark mass at the renormalization point and, via solving the quark DSE, becomes the 'constituent-quark' mass scale at low momenta.

             The second bracket in Eq.~\eqref{vertex:BC} is the transverse part that carries dynamical information from timelike vector-meson poles
             and the quark anomalous magnetic moment.
             Transversality and analyticity demand that  the transverse part must be at least
             linear in the photon momentum $Q$ and vanish at $Q\rightarrow 0$.
             A tensor basis that implements these features automatically was constructed in Ref.~\cite{Kizilersu:1995iz}. It can be written in a compact form~\cite{Eichmann:2012mp}:
              \renewcommand{\arraystretch}{1.4}
             \begin{equation}\label{qpv-newbasis}
             \begin{split}
                 \begin{array}{rl}
                 \tau_1^\mu \!\!\!\!\!\! &= t^{\mu\nu}_{QQ}\,\gamma^\nu\,, \\
                 \tau_2^\mu \!\!\!\!\!\! &= t^{\mu\nu}_{QQ}\,k\!\cdot\! Q\,  \tfrac{i}{2} [\gamma^\nu,\Slash{k}]\,, \\
                 \tau_3^\mu \!\!\!\!\!\! &= \tfrac{i}{2}\,[\gamma^\mu,\slashed{Q}]\,, \\
                 \tau_4^\mu \!\!\!\!\!\! &= \tfrac{1}{6}\,[\gamma^\mu, \slashed{k}, \slashed{Q}]\,,
                 \end{array}\quad
                 \begin{array}{rl}
                 \tau_5^\mu \!\!\!\!\!\! &= t^{\mu\nu}_{QQ}\,ik^\nu\,, \\
                 \tau_6^\mu \!\!\!\!\!\! &= t^{\mu\nu}_{QQ}\,k^\nu \Slash{k}\,, \\
                 \tau_7^\mu \!\!\!\!\!\! &= t^{\mu\nu}_{Qk}\,k\!\cdot\! Q\,\gamma^\nu\,, \\
                 \tau_8^\mu \!\!\!\!\!\! &= t^{\mu\nu}_{Qk}\,\tfrac{i}{2}\,[\gamma^\nu,\Slash{k}]\,,
                 \end{array}
             \end{split}
             \end{equation}
             where $t_{ab}^{\mu\nu} = a\cdot b\,\delta^{\mu\nu} - b^\mu a^\nu$ 
             is transverse to $a^\mu$ and $b^\nu$ and regular in the limits $a,b\rightarrow 0$. 
             We have also employed the antisymmetric combination of three $\gamma-$matrices: $\left[A,B,C\right] = \left[A,B\right] C + \left[B,C\right]A+\left[C,A\right]B$.

            \begin{figure}[t!]
            \centerline{%
            \includegraphics[width=14cm]{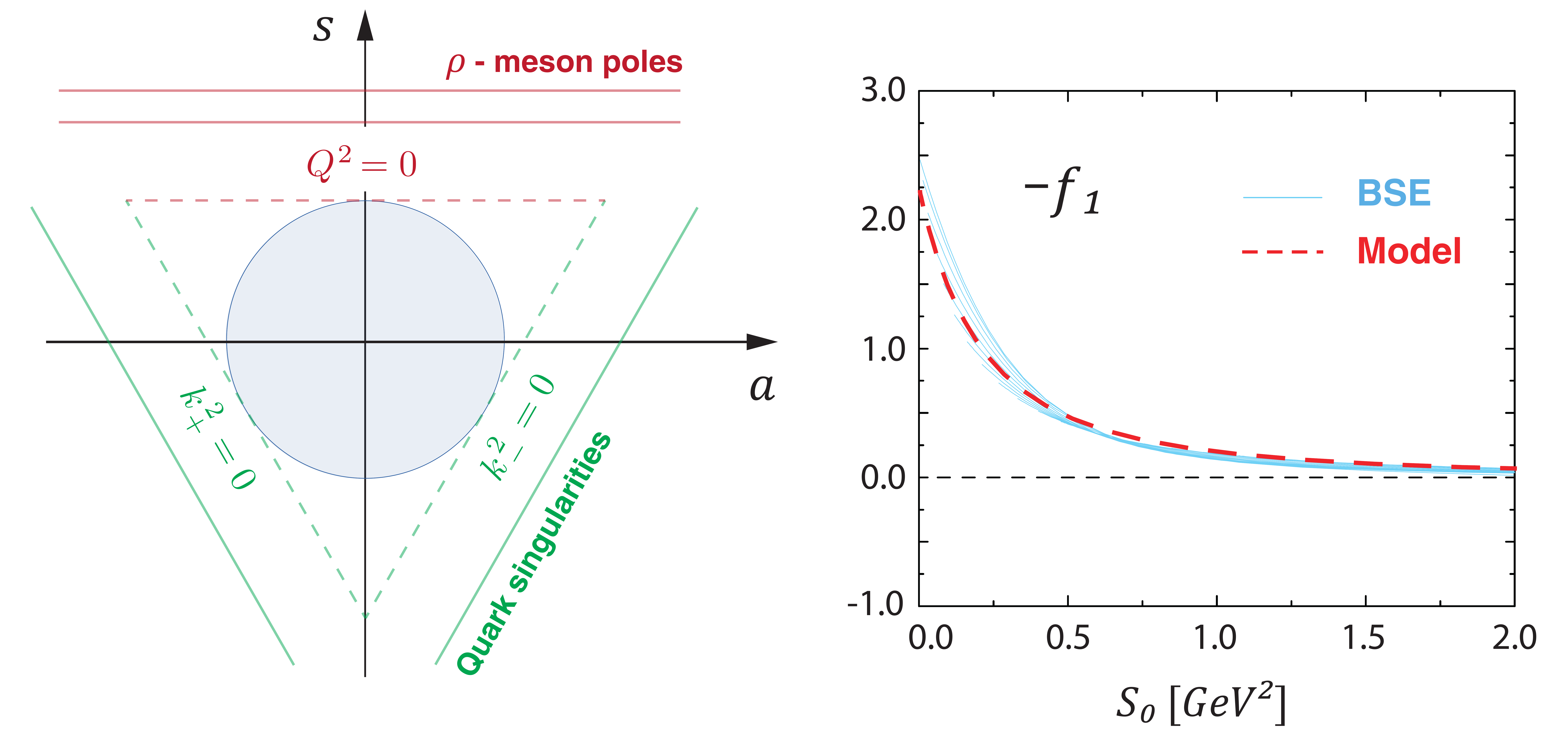}}
            \caption{\textit{Left panel:} Mandelstam plane in the variables $a$ and $s$. \textit{Right panel:} angular dependence of the transverse dressing function $-f_1$.}
            \label{fig:phasespace}
            \end{figure}

             Since the gauge part is already determined by the quark propagator, let us focus on the transverse dressing functions $f_{1\dots 8}(k^2,k\cdot Q,Q^2)$.
             We can express the three Lorentz invariants on which they depend in terms of the variables~\cite{Eichmann:2014xya}
             \begin{equation}
                 \mathcal{S}_0 = \frac{k^2}{3} + \frac{Q^2}{4}\,, \quad
                 a = \frac{k\cdot Q}{\sqrt{3}\,\mathcal{S}_0}\,, \quad
                 s = 1-\frac{Q^2}{2\,\mathcal{S}_0}\,.
             \end{equation}
             The symmetric variable $\mathcal{S}_0$ is a singlet under the permutation group $\mathds{S}^3$ and carries the mass dimension.
             The angular variables $a$ and $s$ form a doublet and constitute the Mandelstam plane which is illustrated in Fig.~\ref{fig:phasespace}.
             The lines of constant $k_+^2$, $k_-^2$ and $Q^2$ are shown together with the spacelike region that forms the interior of a unit circle.
             In the timelike domain one eventually encounters vector-meson poles at $Q^2=-m_\rho^2$, but also quark singularities at timelike values of $k_\pm^2$
             (which may be complex poles or branch cuts instead).
             Charge-conjugation invariance entails that the dressing functions are symmetric
             under a reflection $a\rightarrow -a$.

             Somewhat surprisingly, the transverse dressing functions $f_j$ exhibit only a weak dependence on the variables $a$ and $s$ in the spacelike region.
             This is exemplified in Fig.~\ref{fig:phasespace}:
             the angular dependence of $f_1$ produces only a narrow spread in the symmetric variable $\mathcal{S}_0$.
             At least for crude modeling purposes, one can then parametrize the BSE result by simple multipoles in $\mathcal{S}_0$:
              \renewcommand{\arraystretch}{1.2}
             \begin{equation}\label{transverse-parametrization}
                 f_j(k^2,k\cdot Q,Q^2) \approx \frac{c_j/\Lambda^{n_j}}{(1+\mathcal{S}_0/\Lambda^2)^{n_j}} \qquad \text{for}\quad \begin{array}{l} \mathcal{S}_0 > 0, \\ a^2+s^2 <1\,.\end{array}
             \end{equation}
             The exponents $n_j$ are the dimensions of the basis elements in Eq.~\eqref{qpv-newbasis}.
             The dimensionless coefficients $c_j$ are extracted from our numerical BSE solution and collected in Table~\ref{tab:tab1}.
             The scale $\Lambda=0.65$ GeV yields the closest simultaneous description of all eight dressing functions.

             In Eq.~\eqref{qpv-newbasis} only the elements $\tau_3^\mu$, $\tau_4^\mu$ and $\tau_8^\mu$ are linear in the photon momentum~$Q$ whereas all others depend on higher powers of $Q$.
             Hence, only these can
             contribute to the magnetic moments of hadrons (in addition to the Ball-Chiu vertex).
             Table~\ref{tab:tab1} shows that the $f_3$ component, which encodes the 'anomalous magnetic moment of a quark', is practically zero in rainbow-ladder. 
             The smallness of $f_3$ has the interesting consequence that nucleon magnetic moments calculated from the three-body Faddeev equation
             are generated by the Ball-Chiu vertex alone, whereas the transverse part of the vertex contributes almost nothing~\cite{Eichmann:2011vu}.
             Still, the calculated magnetic moments are reasonably close to their experimental values, with discrepancies believed to be due to pion-cloud effects.
             If a quark anomalous magnetic moment is produced by interactions beyond rainbow-ladder~\cite{Chang:2010hb},
             judging from these results it seems at least unlikely to generate large corrections to form factors.
             Another perhaps more direct test of the properties of the quark-photon vertex is the vector current-current correlator, i.e.,
             the hadronic vacuum polarization which is relevant for the muon $g-2$ puzzle~\cite{Jegerlehner:2009ry,Goecke:2012qm}.

             \begin{table}[t]

         \renewcommand{\arraystretch}{1.0}

                \begin{center}
                \begin{tabular}{  @{\;\;} c @{\;\;\;\;} | c @{\;\;\;\;}c @{\;\;\;\;}c @{\;\;\;\;}c @{\;\;\;\;}c @{\;\;\;\;}c @{\;\;\;\;}c @{\;\;\;\;} c @{\;\;\;\;}     }

                            & $f_1$    & $f_2$    & $f_3$    & $f_4$    & $f_5$    & $f_6$    & $f_7$    & $f_8$        \\   \noalign{\smallskip}\hline\noalign{\smallskip}

                    $c_j$   & $-1.0$   & $0.15$    & $0$      & $1.0$   & $-1.3$   & $-0.3$   & $0.5$   & $-0.3$      \\
                    $n_j$   & $2$      & $5$      & $1$      & $2$      & $3$      & $4$      & $4$      & $3$

                \end{tabular}
                \end{center}

               \caption{Simple parametrization of the rainbow-ladder BSE result for the quark-photon vertex, cf.~Eq.~\eqref{transverse-parametrization}.
                        The values correspond to the standard (central) parameter set for the quark-gluon interaction in Ref.~\cite{Maris:1999nt}.}
               \label{tab:tab1}

        \end{table}

            \begin{figure}[t!]
            \centerline{%
            \includegraphics[width=13.5cm]{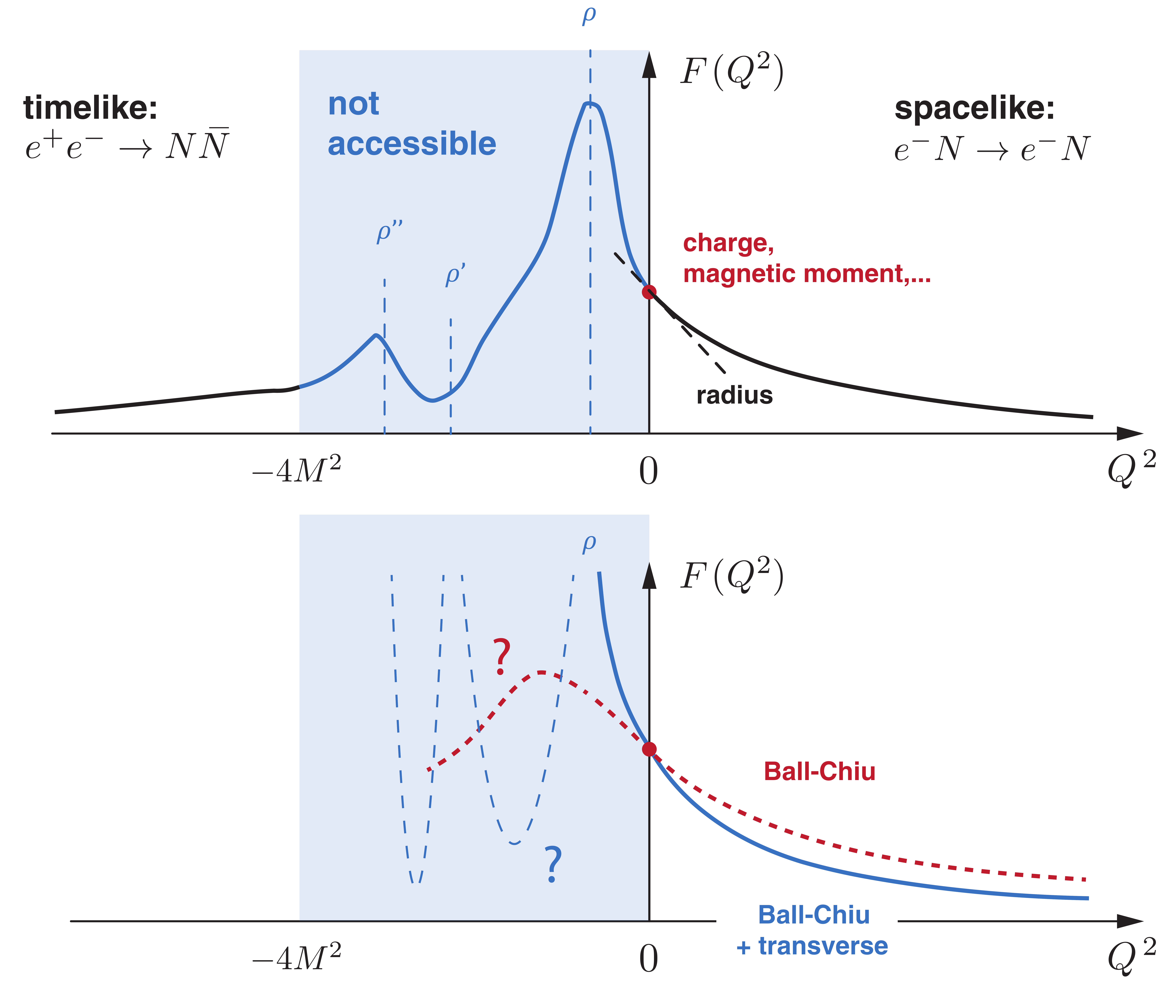}}
            \caption{Sketch of a generic form factor in the spacelike and timelike region.
                     The upper plot exemplifies a typical 'experimental curve' with characteristic $\rho-$meson bumps in the timelike region, modeled after those in the pion form factor.
                     The lower plot illustrates a typical calculated result in the spacelike region with its separation into Ball-Chiu and transverse part. Only the latter
                     contains $\rho-$meson poles. As long as the truncation does not dynamically accommodate a $\rho\rightarrow\pi\pi$ decay, these poles do not carry widths.}
            \label{fig:timelike-ffs}
            \end{figure}

             Finally, the structure of the quark-photon vertex is reflected in the timelike properties of hadrons, as it inherits the vector-meson pole structure from the quark four-point function.
             'Vector-meson dominance' is thus a self-consistent outcome of the inhomogeneous BSE, as explicitly demonstrated in Ref.~\cite{Maris:1999bh} and illustrated in Fig.~\ref{fig:timelike-ffs}:
             vector-meson poles are dynamically generated at the quark-gluon level.
             However, vector-meson dominance is only part of the full picture. The Ball-Chiu vertex depends only on the quark propagator and cannot produce timelike vector-meson poles.
             Nevertheless it is the dominant contribution in the spacelike region: it reproduces the nucleon's charge at $Q^2=0$ and, effectively, also its magnetic moments.
             Timelike poles can only come from the transverse piece in Eq.~\eqref{vertex:BC}, which does not contribute to the charge and vanishes for $Q^2 \rightarrow\infty$.
             With this perspective, gaining a deeper understanding of form factors from vector-meson dominance formulas alone seems a bit too optimistic.

             On the other hand, it is the \textit{same} quark-photon vertex that enters in pion and nucleon form-factor calculations. Since the vertex alone
             carries the resonance dynamics, it is conceivable that the resonance structure in the nucleon's unphysical window,
             below $N\bar{N}$ threshold, is similar to what is known from the pion form factor, as indicated in Fig.~\ref{fig:timelike-ffs}. This yields a straightforward prescription, for example,
             for modeling the $\Delta\to N e^+ e^-$ Dalitz decay in $pp$ scattering~\cite{Buss:2011mx,Ramalho:2012ng}.
             In any case, as long as the truncation that is employed in the vertex BSE does not dynamically implement a $\rho\to\pi\pi$ decay, the calculated poles are real and do not carry widths: 
             in rainbow-ladder, hadrons are stable bound states that do not decay. 
             Hence, in order to describe the timelike properties of hadrons, it will be crucial to develop interactions beyond rainbow-ladder that accommodate such features.

     \begin{figure*}[t]
     \center{
     \includegraphics[width=13.5cm]{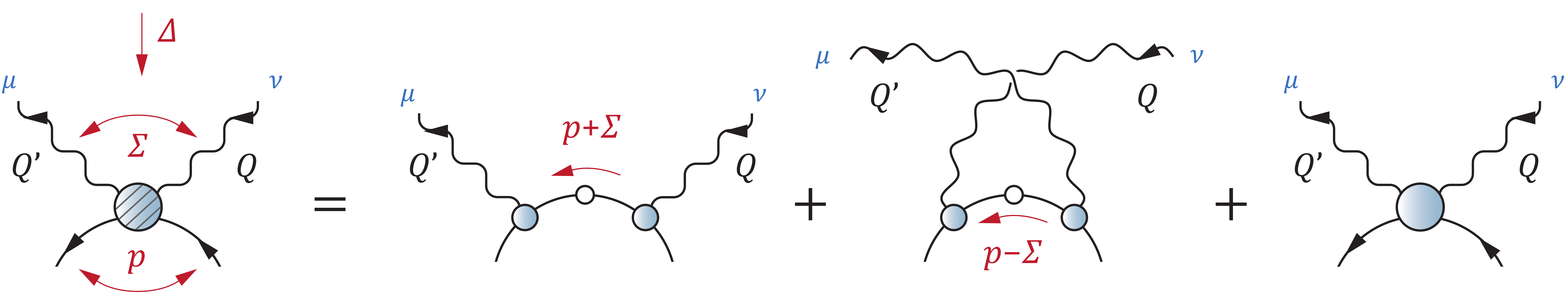}}
        \caption{Separation of the quark Compton vertex into Born terms and a 1PI part. }
        \label{fig:qcv-born}
     \end{figure*}

       \section{Quark Compton vertex}

             While form factors probe certain aspects of the electromagnetic structure of hadrons,
             two-photon processes present a far richer spectrum of applications: from nucleon polarizabilities
             to nucleon structure functions, generalized parton distributions or the proton radius puzzle. These are all, in some way or another,
             related to the nucleon's Compton scattering amplitude.
             In the same way as electromagnetic form factors test the underlying quark-photon vertex, the Compton amplitude
             depends upon the quark two-photon or quark Compton vertex~\cite{Eichmann:2012mp,Eichmann:2013bd}.

             The quark Compton vertex can be written as the sum of Born terms plus a one-particle-irreducible structure part, cf.~Fig.~\ref{fig:qcv-born},
             and mirrors the properties of the quark-photon vertex in many respects.
             Similarly as in Fig.~\ref{fig:qpv}, it can be defined as the contraction of the $q\bar{q}$ Green function
             with the Born terms. Hadronic states in the Green function appear as $t-$channel poles in the Compton vertex (and
             nucleon Compton amplitude): pion, scalar, axial-vector poles etc.
             One can derive an inhomogeneous BSE for the vertex that depends again upon the $q\bar{q}$ kernel.
             Its rainbow-ladder solution, together with the $t-$channel poles it produces, was presented in Ref.~\cite{Eichmann:2012mp}.

             The Compton vertex satisfies a WTI which allows to write it like Eq.~\eqref{vertex:BC}, as the sum of a 'gauge part' and a
                   purely transverse piece.
             The transverse part consists of 72 tensor basis elements;
                   transversality and analyticity imply that they must vanish with at least two powers in the photon momenta. Applied to the (onshell)
                   nucleon Compton amplitude, this is the low-energy theorem in Compton scattering, and the photon momentum counting
                   is the same as the counting in chiral perturbation theory.
             One can then identify the presumably dominant tensor structures in the vertex: for example,
             \begin{equation}
                 t^{\mu\alpha}_{Q'p}\,t^{\alpha\nu}_{pQ} \quad \text{and} \quad \varepsilon^{\mu\alpha}_{Q'p}\,\varepsilon^{\alpha\nu}_{pQ}\,, \qquad \text{with} \quad
                 \varepsilon^{\mu\nu}_{ab} = \varepsilon^{\mu\nu\alpha\beta}\,a^\alpha\,b^\beta\,,
             \end{equation}
             encode the 'quark electric and magnetic polarizabilities', respectively.

             In contrast to the form factor example, the quark Compton vertex does not generate the full nucleon Compton scattering amplitude but only the subset of its 'handbag' diagrams.
             Those alone are not sufficient for electromagnetic gauge invariance~\cite{Eichmann:2012mp};
             nevertheless, insight can be gained from analyzing the individual strengths of the vertex contributions and their importance at the hadron level.
             A more direct test of the Compton vertex is the photon four-point function where gauge invariance is satisfied without the need for additional terms~\cite{Eichmann:2011ec,Goecke:2012qm}.
             The four-point function enters in the light-by-light contribution to the muon anomalous magnetic moment and, to some extent,
             provides an 'experimental probe' of the structure of the Compton vertex --- much like form factors probe the properties of the quark-photon vertex.
             Work in this direction is currently in progress.

             \bigskip

       \textbf{Acknowledgments.} I am grateful to C. S. Fischer and R. Williams for discussions. This work was supported by the German Science Foundation (DFG)
             under project number DFG TR-16, and by the Austrian Science Fund (FWF) under project numbers J3039-N16 and P25121-N27.

\end{document}